\documentclass[12pt]{article}
\usepackage{graphicx}
\usepackage{tabularx}
\usepackage{a4}
\usepackage{url}
\usepackage{longtable, lscape}
\usepackage[usenames, dvipsnames]{color}
 \usepackage{pdflscape}

\usepackage{multirow, makecell, bigstrut, booktabs, float}
\usepackage{array}

\usepackage{setspace}
\usepackage[fleqn]{amsmath}
\usepackage{amsfonts}
\usepackage{amssymb}
\usepackage{subfigure}
\usepackage{tcolorbox}
\usepackage{mathrsfs}  
\usepackage{bbm}

\usepackage{adjustbox}
\usepackage{nicefrac}
\usepackage{pgfplots}
\usepackage{authblk}

\graphicspath{{Bilder/}{Bilder/OriginalPlots/}{Bilder(formerWeibullCode)/}}

\begin{document}

\title{The population-attributable fraction for time-dependent exposures and competing risks -- A discussion on estimands}

\author[1,2]{Maja von Cube\thanks{cube@imbi.uni-freiburg.de, Ernst-Zermelo Strasse 1, 79104 Freiburg}}

\author[1,2]{Martin Schumacher}

\author[3,4]{S\'{e}bastien Bailly}
\author[5,6]{Jean-Fran\c{c}ois Timsit}
\author[8,9]{Alain Lepape}
\author[7,9]{Anne Savey}
\author[7]{Anais Machut}
\author[1,2]{Martin Wolkewitz}

\affil[1]{Institute of Medical Biometry and Statistics, Faculty of Medicine and Medical Center - University of Freiburg, Freiburg, Germany}

\affil[2]{Freiburg Center for Data Analysis and Modelling, University of Freiburg, Freiburg, Germany}

\affil[3]{HP2 laboratory, University of Grenoble Alpes, Grenoble, France}

\affil[4]{Department of Physiology and Sleep, Grenoble Alpes University Hospital, Grenoble, France}

\affil[5]{UMR 1137 IAME Inserm/Universite Paris Diderot, Paris, France}

\affil[6]{APHP Medical and Infectious Diseases ICU, Bichat Hospital, Paris, France}

\affil[7]{CPIAS Auvergne-Rhone-Alpes, Hospices Civils de Lyon, Lyon, France}

\affil[8]{Clinical research Unit, Critical care, Lyon Sud University Hospital, Hospices Civils de Lyon, Lyon, France}

\affil[9]{Laboratory of Emerging Pathogens, International Center for Infectiology Research (CIRI), Inserm U1111, CNRS UMR5308, ENS de Lyon, UCBL1, Lyon, France}

\renewcommand\Authands{ and }

\maketitle

\section*{Abstract}

The population-attributable fraction (PAF) quantifies the public health impact of a harmful exposure. Despite being a measure of significant importance an estimand accommodating complicated time-to-event data is not clearly defined. We discuss current estimands of the PAF used to quantify the public health impact of an internal time-dependent exposure for data subject to competing outcomes. To overcome some limitations, we proposed a novel estimand which is based on dynamic prediction by landmarking. In a profound simulation study, we discuss interpretation and performance of the various estimands and their estimators. The methods are applied to a large french database to estimate the health impact of ventilator-associated pneumonia for patients in intensive care.
\medskip

\textbf{Keywords: Population-attributable risk, time-dependent exposure, competing risks, hospital-acquired infection, mortality}

\section{Introduction}

An important intention of public health decisions is the containment of fatal exposures such as infectious diseases. A recent example is the fear of
an increasing spread of antimicrobial resistance \cite{laxminarayan2013antibiotic}. The need for action is based on the number of lives that could be spared if nosocomial infections (NIs) could be prevented or entirely cured. Thus, quantifying the threat of an exposure for a population is of main interest when
taking decisions about prevention programs and the development of new drugs.

To quantify the public health impact of a harmful exposure Levin \cite{Levin1953PAF} defined the population-attributable fraction (PAF). It expresses the fraction of all cases that are attributable to the exposure and is often interpreted as percentage of preventable cases.
The PAF is commonly estimated as a static measure over a specific time period. A generalization has been proposed for time-to-event data \cite{chen2006attributable, chen2010attributable, samuelsen2008attributable, laaksonen2010piecewise,laaksonen2010estimation, sjolander2016cautionary, zhao2017onestimation, sjolander2017doubly}. However, these approaches are not appropriate for data settings with an internal time-dependent exposure. We emphasize that harmful exposures are often naturally time-dependent since -- unlike in randomized clinical trials -- the time of onset cannot be chosen by the researcher. Moreover, the outcome is often unobservable due to competing risks. This is the case if attributable risk is quantified by cause-specific mortality or non-mortal endpoints. For example, in hospital epidemiology, the burden of a health-care associated infection (i.e. an NI) is often quantified in terms of death in the hospital or intensive care unit (ICU). Then, even though observational cohort studies in ICUs have often complete follow-up, one has to account for discharge alive as a competing risk to death in the ICU \cite{wolkewitz2014interpreting}. Additionally, infections occur over the course of time. Ignoring the fact that NI is a time-dependent exposure leads to the so-called time-dependent bias \cite{schumacher2013hospital}. Another challenging aspect in this data situation is the adjustment for time-varying confounding which is essential to draw conclusions from observational studies. Thus, major difficulties in defining and estimating the PAF arise when the exposure is time-dependent.

Literature on the PAF for time-dependent exposures and competing risks is sparse \cite{sjolander2017doubly} and inconsistent. Different proposed estimands \cite{schumacher2007attributable, bekaert2010adjusting} lead to different conclusions \cite{vonCube2019beweis}. In \cite{vonCube2019beweis}, we study the differences of the approaches by Schumacher et al. \cite{schumacher2007attributable} and Bekaert et al. \cite{bekaert2010adjusting}. In \cite{vonCube2019lm}, we propose a novel estimand of the PAF which is based on dynamic prediction by landmarking \cite{van2011dynamic}.
The novel approach overcomes some limitations of interpretation and estimation of the current approaches by Schumacher et al. \cite{schumacher2007attributable} and Bekaert et al. \cite{bekaert2010adjusting}. It provides clinically relevant implications and the estimator is easily adjustable for time-dependent confounders.

The purpose of this article is to introduce the various ways of defining the PAF for complex time-to-event data and to draw attention to the differences with regard to interpretation. Our investigations include also a study of basic aspects of the corresponding estimators of the PAF. In Section 2, we introduce the established approaches used to define and estimate the PAF in hospital epidemiology. Moreover, the novel approach which is discussed in detail in \cite{vonCube2019lm} is outlined. A discussion and comparison of the various estimands and their estimators based on a simulation study are presented in Section 3. In Section 4, we analyse a large sample of ventilated patients in intensive care to estimate attributable mortality and the percentage of preventable ICU death cases if ventilator-associated pneumonia (VAP) was avoided. For simplicity, all the methods are explained and discussed based on the example of attributable ICU death cases due to NIs. However, they are applicable to any other survival data setting with binary time-dependent exposure and competing risks. Even though our focus is on data settings with complete follow-up, all estimands can be estimated with censored time-to-event data.

\section{The population-attributable fraction for time-dependent exposures: Estimands and estimators}

\subsection{Initial definition of the PAF}

The PAF is defined by
\begin{equation}
\textrm{PAF}=\frac{P(D=1)-P(D=1|E=0)}{P(D=1)},
\label{PAFl}
\end{equation}
where $D$ is the random variable of a dichotomous outcome and $E$ of a dichotomous exposure \cite{benichou2001review}. The sampling schemes that correspond to such an estimand are usually cross-sectional studies and cohort studies of fixed length. Note that $D$ and $E$ are both observable random variables. 
An equivalent definition in terms of the relative risk (RR) \cite{miettinen1974proportion} is
\begin{equation}
\textrm{PAF}=P(E=1|D=1)\times \frac{RR-1}{RR}.
\label{PAFm}
\end{equation} 
This representation shows that the PAF takes both the prevalence of the exposure and the strength of association of exposure and outcome into account. In cross-sectional studies and cohort studies of fixed length with a time-independent exposure the PAF can be interpreted causally if it has been sufficiently adjusted for confounding.
Then, it can be interpreted as proportion of preventable death cases if exposure was completely extinct \cite{benichou2005attributable}.

\subsubsection{Defining $\textrm{PAF}_{crude}$}

If follow-up is complete, information on exposure and outcome is often summarized in a fourfold table. A patient that is discharged alive has the realization $D=0$. A patient that died in the ICU has the realization $D=1$. Patients acquiring an infection during their ICU stay have $E=1$, otherwise $E=0$. Then, the percentage of attributable ICU death cases is often accessed with \eqref{PAFl} or \eqref{PAFm}. 
The resulting estimand relates the proportion of patients that were ever infected to the proportion of patients that remained uninfected until the end of their ICU stay. We denote this estimand with $\textrm{PAF}_{crude}$. In clinical literature this is still the most commonly used approach. However, $\textrm{PAF}_{crude}$ ignores the time-dependencies of exposure and outcome and is therefore only a crude measure of attributable risk.

\subsubsection{Estimating $\textrm{PAF}_{crude}$}

Estimation of $\textrm{PAF}_{crude}$ can be performed with the R-package 'AF' \cite{dahlqwist2016model} which is based on generalized-linear models (GLMs). The R-package allows for adjustment for confounding based on the maximum likelihood estimator from a logistic model as proposed by Greenland and Drescher \cite{greenland1993maximum}. Note that in case of censoring, $\textrm{PAF}_{crude}$ can be estimated by $\textrm{PAF}_o(\tau)$, where $\tau$ denotes the end of follow-up. This approach is explained in the following Section 2.2 and discussed in more detail in Section 3.

\subsection{Defining $\textrm{PAF}_o(t)$}

In order to account for the timing of exposure and outcome, Schumacher et al. \cite{schumacher2007attributable} extended the definition of $\textrm{PAF}_{crude}$ to
\begin{equation}
\textrm{PAF}_o(t)=\frac{P(D(t)=1)-P(D(t)=1|E(t)=0)}{P(D(t)=1)},
\label{eqPAF_S}
\end{equation}
where $D(t)$ is the random variable that indicates if the patient died in the ICU by time $t$ ($D(t)=1$ if the patient died) and $E(t)$ if the patient acquired an NI by $t$ ($E(t)=1$ if the patient acquired an NI within $(0,t]$). Both $E(t)$ and $D(t)$ are observable random variables. By definition of NIs patients are naturally unexposed at study entry. Thus, $E(0)=0$ for all patients. $\textrm{PAF}_o(t)$ relates the proportions of patients that died infection-free within $(0,t]$ to the proportion of all patients that died. Therefore, it is interpretable as the \textit{observable} proportion of attributable death cases until time $t$ among all initially admitted patients \cite{vonCube2019beweis}.

\subsubsection{Estimating $\textrm{PAF}_o(t)$}

\begin{figure*}
\centering
\includegraphics[width=\textwidth]{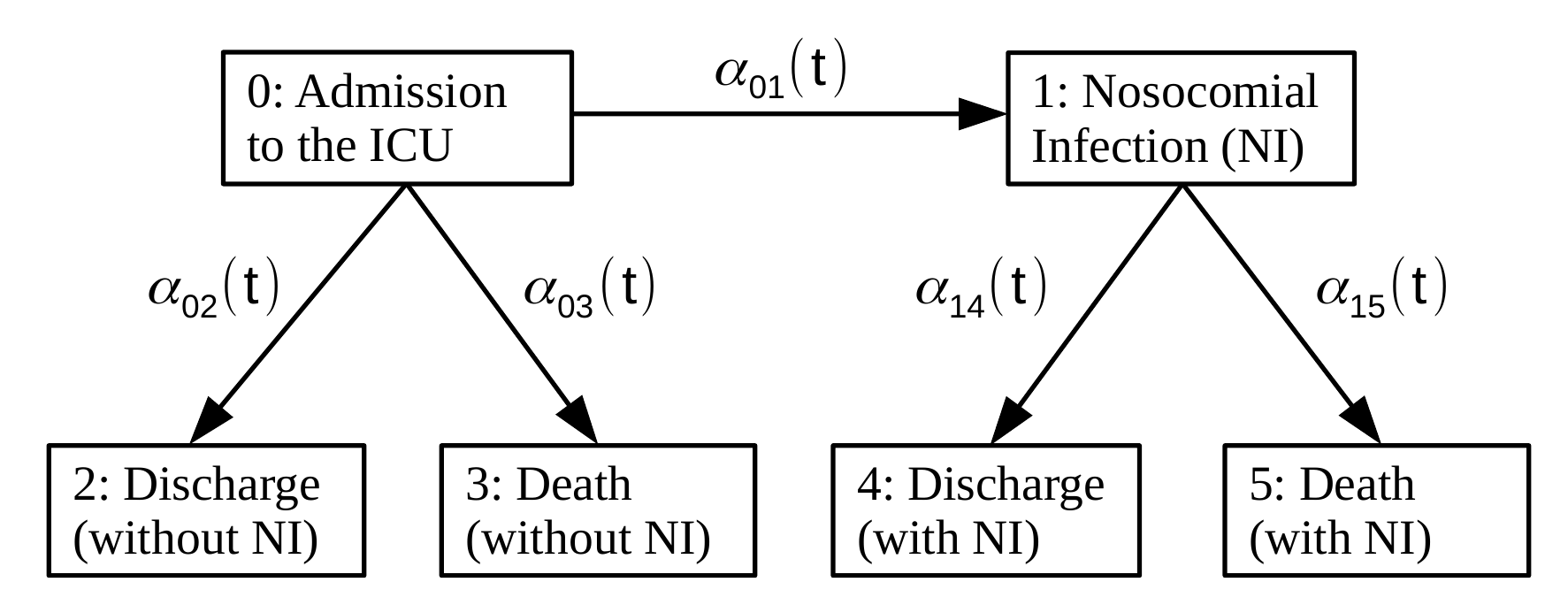}
 \caption{Extended Illness-Death Model with hazard rates $\alpha_{01}(t)$, $\alpha_{02}(t)$, $\alpha_{03}(t)$, $\alpha_{14}(t)$ and $\alpha_{15}(t)$.}
  \label{fig:extendedIllnessDeath}
\end{figure*}

For identification and estimation Schumacher et al. \cite{schumacher2007attributable} proposed to use the multi-state model shown in Figure \ref{fig:extendedIllnessDeath}. This model is often called 'extended illness-death model' \cite{beyersmann2011competing}. The exposure is modelled as an intermediate state (State 1), the outcome death in the ICU as State 3 if the patient died unexposed and as State 5 if the patient acquired an NI before death. Analogously, discharge alive is modelled as State 2 and State 4.

Based on the extended illness-death model and with the Bayes' Theorem the conditional probability of ICU death is identifiable by
\begin{equation}
P(D(t)=1|E(t)=0)=\frac{P_{03}(0,t)}{P_{00}(0,t)+P_{02}(0,t)+P_{03}(0,t)},
\label{condP03}
\end{equation}
where $P_{0j}(0,t)$ ($j=0,2,3$) denote the transition probabilities of the extended illness-death model \cite{schumacher2007attributable}.
Moreover, the sum of the risk to die without and with the exposure results in an estimand of overall mortality:
\begin{equation}
P(D(t)=1)=P_{03}(0,t)+P_{05}(0,t).
\label{P_Dt}
\end{equation}

Estimation of the transition probabilities can be, for example, performed with the Aalen-Johansen estimator. To adjust for confounding, \cite{coeurjolly2012attributable} proposed a semi-parametric modelling approach. Alternatively, the Aalen-Johansen estimators can be estimated within strata based on baseline covariates. Confidence-intervals can be obtained with a bootstrap approach.

\subsection{Defining $\textrm{PAF}_c(t)$}

An alternative definition of the PAF for data settings with time-dependent exposure and competing risks was motivated by Bekaert et al. \cite{bekaert2010adjusting} and formalized by von Cube et al. \cite{vonCube2019beweis}. We denoted this estimand with $\textrm{PAF}_c(t)$. The initial definition was based on counterfactual outcomes. In this article, we use a slightly different approach to define $\textrm{PAF}_c(t)$. 
This approach is based on the definition of the PAF (for cross-sectional studies and cohort studies of fixed length) by Eide et al.\cite{eide2001attributable}.

Let $D(t)$ be, as defined above, the observable random variable of death by time $t$ and $P$ the corresponding distribution in the target population. Thus, $P(D(t)=1)$ is - as in \eqref{eqPAF_S} - the (observable) overall death risk. To denote the hypothetical death risk of this population if the exposure could be eliminated - all other things left equal - we define the distribution $P_0$ of $D(t)$.
Then, $\textrm{PAF}_c(t)$ is given by
\begin{equation}
\textrm{PAF}_c(t)=\frac{P(D(t)=1)-P_0(D(t)=1)}{P(D(t)=1)},
\label{eqPAF_B}
\end{equation}
where $P_0(D(t)=1)$ is the hypothetical death risk at $t$ had all patients remained unexposed. 
$\textrm{PAF}_c(t)$ is interpreted as the percentage of preventable ICU death cases over the course of time had all patients remained infection-free.

\subsubsection{Estimating $\textrm{PAF}_c(t)$}

To identify $\textrm{PAF}_c(t)$, we must identify the two distributions, $P$ and $P_0$, of the random variable $D(t)$. 
$P$ corresponds to the distribution of $D(t)$ in the observable population and can be identified by \eqref{P_Dt}. $P_0$ is the hypothetical distribution after a manipulation of the transition intensities of the extended illness-death model (Figure \ref{fig:extendedIllnessDeath}). We only consider an indirect manipulation of the death risk by setting the infection hazard to zero. The resulting distribution of $D(t)$ corresponds to $P_0$. Now $P_0(D(t)=1)$ can be identified with the cumulative-incidence function (CIF) of death without NI. For estimation, patients that acquire an infection are treated as censored observations instead of accounting for NI as a competing risk. An estimator for the CIF is, for example, the Aalen-Johansen estimator. We denote this estimator with $\hat{P}_{03_0}(0,t)$. For details we refer to \cite{vonCube2019beweis} and the Appendix. 

Bekaert et al. propose two estimators of $P_0(D(t)=1)$. The first is called naive, as it is not adjusted for counfounding factors. We emphasize that this naive estimator is an estimator for the conditional probability function \eqref{condP03} \cite{vonCube2019beweis}. Thus, the resulting estimand of the PAF is $\textrm{PAF}_o(t)$.
Furthermore, Bekaert et al. propose to adjust $\textrm{PAF}_c(t)$ by adjusting $P_0(D(t)=1)$ using inverse probability weights (IPWs). The weights denote the probability of being uninfected at time $t$ conditional on observable covariates (time-dependent and time-independent). The weights can be derived using pooled logistic regression. The causal interpretation is justified if the IPWs are correctly specified at each time $t$ and under the usual assumptions of marginal structural models \cite{robins2000marginal}. We emphasize that in the absence of confounding, this estimator is equivalent to $\hat{P}_{03_0}(0,t)$ \cite{vonCube2019beweis}. The resulting estimand is $\textrm{PAF}_c(t)$. In this article, we are mainly interested in the estimands of the PAF and consider simulation studies without confounding. However, in practice adjustment for time-varying confounding is essential to draw causal conclusions. Then, the IPW-approach by Bekaert et al. \cite{bekaert2010adjusting} provides a way to adjust $\widehat{\textrm{PAF}}_c(t)$ for time-varying confounding. To obtain confidence intervals a bootstrap approach can be used.

\subsection{Defining $\textrm{PAF}_{LM}$}

\begin{figure*}
\centering
\includegraphics[scale=0.8]{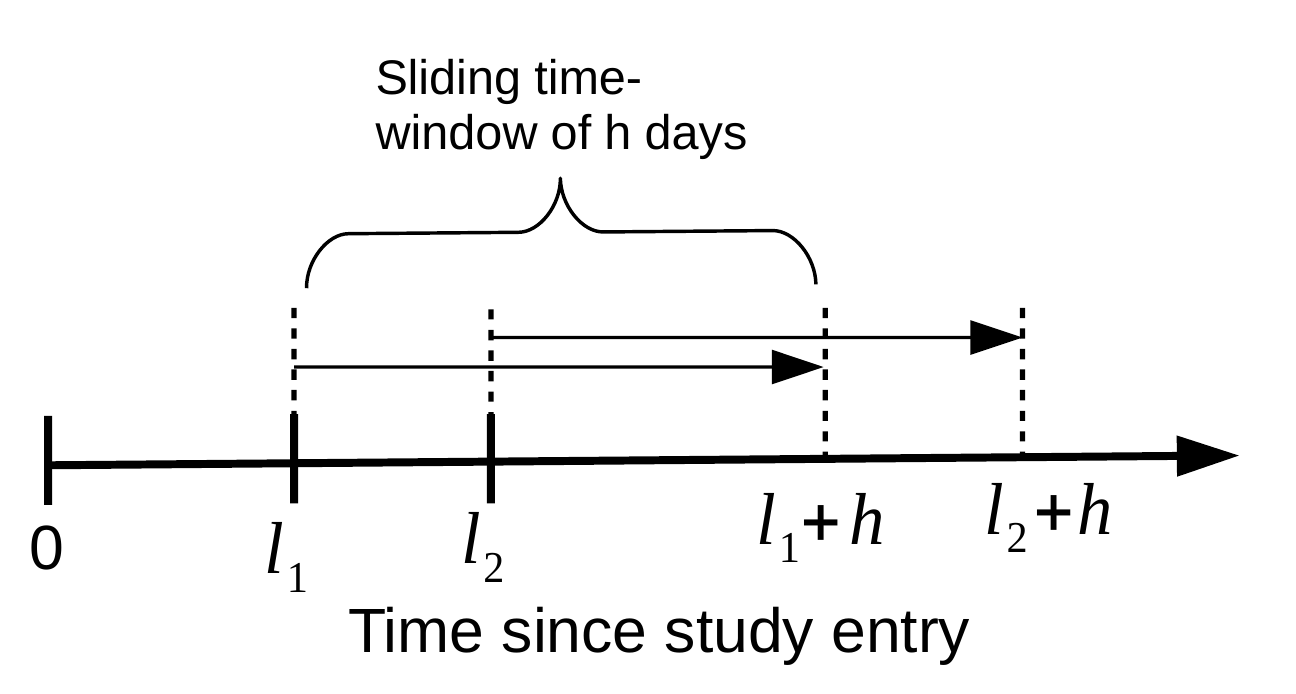}
\caption{Dynamic prediction by landmarking: $l_1$, $l_2$ are landmarks, $h$ is the time window of fixed length; $PAF(l_1, h)$ is a summary measure over the time-interval $(l_1, l_1+h]$.}
\label{fig:LM}
\end{figure*}

A different way to define the PAF while accounting for the time-dependencies of exposure and outcome is via dynamic prediction by landmarking \cite{vonCube2019lm}. Generally speaking, this is done by choosing a set of relevant time points on the study time scale (landmarks) and an adequate time window (e.g. the mean length of stay in the ICU). At each landmark the exposure state is updated and kept fixed over the time-window.
Then, at each landmark a PAF within the time window can be defined with the time-independent definitions \eqref{PAFl} or \eqref{PAFm}. The PAF at a specific landmark provides information for the patient population that remained in the ICU until that day. The exposed patients are those patients, that are still in the ICU at the landmark and acquired an infection some time before or at the landmark. The unexposed patients are those that are NI-free at the landmark. As the PAF is defined over a specific time window it quantifies the proportion of attributable death cases which occur within the time window.
These attributable cases could be prevented if NIs could be eliminated for patients being infected at the landmark if the Markov assumption holds \cite{vonCube2019lm}. By considering a whole set of landmarks, we account for the time-dependency of infections.
A graphical presentation of $\textrm{PAF}_{LM}$ is given in Figure \ref{fig:LM}.

Let $A_l$ indicate whether the patient is still at risk at landmark $l$, $E_l$ if the patient was exposed at $l$ and $D_{l,h}$ if the patient died within $(l,l+h]$. Then, the PAF at a landmark $l$ within a time window $(l,l+h]$ is defined by
\begin{equation}
\textrm{PAF}(l,h)=\frac{P(D_{l,h}=1|A_l=1)-P(D_{l,h}=1|E_l=0,A_l=1)}{P(D_{l,h}=1|A_l=1)}
\end{equation}
or equivalently as
\begin{align}
\textrm{PAF}(l,h)&=P(E_l=1|A_l=1, D_{l,h}=1)\times \frac{RR_{l,h}-1}{RR_{l,h}}\nonumber\\
 &=P_{E_l}\times \frac{RR_{l,h}-1}{RR_{l,h}}
 \label{PAF(s,t)_def}
\end{align}
where $P_{E_l}$ is the prevalence of infection at time $l$ among death cases occurring within the time window $(l,l+h]$ and $RR_{l,h}$ is the RR of death within $(l,l+h]$ depending on the infection state at time point $l$. $RR_{l,h}$ is formally defined by
\begin{equation}
\textrm{RR}_{l,h}=\frac{P(D_{l,h}=1|A_l=1,E_l=1)}{P(D_{l,h}=1|A_l=1, E_l=0)},
\end{equation}
$\textrm{PAF}_{LM}$ is the set of estimands $\textrm{PAF}(l,h)$s over all landmarks. It accounts for the time-dynamics of the study population by considering a range of time points $l$. Remark that at each landmark we consider a different target population.

\subsubsection{Estimating $\textrm{PAF}_{LM}$}

A major strength of $\textrm{PAF}_{LM}$ is the simplicity of its' estimation. Generally, time-dependent confounding is a major challenge in the estimation of the PAF for time-dependent exposures. However, at a landmark $l$ estimation of $\textrm{PAF}_{LM}$ is equivalent to estimation of the PAF in a data setting with baseline exposure and fixed length of follow-up. 
More explicitly, estimation of $\textrm{PAF}(l,h)$ is based on estimation methods for $\textrm{PAF}_{crude}$ since at each landmark and for every chosen time window, information of exposure and outcome can be summarized in a fourfold table. Time-varying confounding is accounted for by updating the covariate values at every landmark.

For estimation, we must first identify a set of relevant landmarks. The choice of landmarks depends on the number of exposed and unexposed patients in the study sample. The number of patients in each group must be sufficiently large such that the regression model converges. Inference at a landmark $l$ is performed on a so-called landmark dataset \cite{van2011dynamic}. For a specific landmark $l$, it includes information on the patients that are still at risk at $l$. All other patients are excluded. The necessary information comprise exposure state at $l$, outcome state at $l+h$ and covariate values at $l$.
If the patient is still alive at $l+h$ or experienced a competing event, then $D_{l,h}=0$ otherwise $D_{l,h}=1$. In case of censoring pseudo-values for $D_{l,h}$ as proposed by Nicolaie et al. \cite{nicolaie2013dynamic} can be used \cite{vonCube2019lm}. 
With the landmark dataset at landmark $l$, $\textrm{PAF}(l,h)$ can be estimated with any method available for the time-independent PAF (e.g. \cite{greenland1987variance, greenland1993maximum, sjolander2011estimation}). Confidence intervals can be obtained in the same way.

Finally, van Houweling and Putter \cite{van2011dynamic} proposed to smooth the separate estimates $\widehat{\textrm{PAF}}(l,h)$ to increase the efficiency of the estimators.
In principle, two different approaches can be used to obtain a smooth curve of the separate $\textrm{PAF}(l,h)$s over all landmarks. Firstly, smoothing methods like splines in a linear model or local smoothers such as "loess" can be applied on the separately estimated $\widehat{\textrm{PAF}}(l,h)$. 
Secondly, a so-called supermodel can be used to obtain a smooth curve over all landmarks directly without first fitting the separate models \cite{van2011dynamic}. The supermodel is basically pooled logistic regression on the landmark datasets stacked together to one large dataset. A regression model is fitted by accounting for possible time-varying effects at the landmarks via interaction terms. For more details, we refer to \cite{vonCube2019lm}.

\section{Comparison of the estimands and estimators}

\subsection{General aspects on the estimands}

The estimands $\textrm{PAF}_o(t)$ and $\textrm{PAF}_c(t)$ are directly comparable, since they are both cumulative measures of attributable risk over the course of time. The target population of these two estimands are all patients admitted to the ICU or more generally all patients initially entering the study. The $\textrm{PAF}_{crude}$ can be viewed as a summary measure of $\textrm{PAF}_o(t)$ as - in the absence of censoring - we have $\textrm{PAF}_o(\tau)=\textrm{PAF}_{crude}$.

$\textrm{PAF}_o(t)$ and $\textrm{PAF}_c(t)$ differ in the choice of unexposed patients. In the definition of $\textrm{PAF}_o(t)$ the unexposed patients are those patients that did not acquire an infection until $t$. Thus, they are an \textit{observable} time-varying \textit{subpopulation} of the target population. In contrast, the unexposed patients considered in $\textrm{PAF}_c(t)$ are a \textit{hypothetical} patient population that differs from the factual one in being infection-free. As a consequence, in the definition of $\textrm{PAF}_c(t)$ the same population is compared under two distinct conditions.

The $\textrm{PAF}_{LM}$ differs from $\textrm{PAF}_o(t)$ and $\textrm{PAF}_c(t)$ in many aspects. Firstly, it is not a cumulative measure over the course of time. Instead, similarly to $\textrm{PAF}_{crude}$, it summarizes the information within a specific time window. Secondly, $\textrm{PAF}_{LM}$ conditions on patients being still at risk at certain time points (the landmarks). This means that - in contrast to $\textrm{PAF}_{crude}$, $\textrm{PAF}_o(t)$ and $\textrm{PAF}_c(t)$ - the \textit{target population} varies with time. The $\textrm{PAF}_{LM}$ is a summary measure of population-attributable risk of patients still in the ICU after a certain amount of days. In Table 2, we present the main characteristics of the four estimands and corresponding estimators of the PAF. The simulation study demonstrates not only the behaviour of the estimators but also the different interpretation of the estimands under ideal conditions (i.e. no confounding).

\subsection{Performance in a simulation study}

All the proposed estimands of the PAF can be identified with the transition probabilities of the extended illness-death model (Figure \ref{fig:extendedIllnessDeath}), which are determined by the five cause-specific hazard rates. Therefore, data generation was based on the extended illness-death model. The data was generated and evaluated with the statistical software R and the R code for the data simulation is an extension of the simulation code provided by Heggland et al. \cite{heggland2015estimating}. We assumed constant hazards and more generally time-varying Weibull hazards to obtain different data situations. The effect of the time-dependent exposure on the outcome of interest (death in ICU) was modelled either directly via an increased death hazard after an infection or indirectly via a decreased discharge hazard after an infection. The parameters of the various scenarios are presented in Table \ref{tab:scenarios}.

Each scenario was simulated 100 times with a sample size of 10,000 or 2,000 observations. In each run, we obtained estimates of $\textrm{PAF}_{crude}$, $\textrm{PAF}_o(t)$, $\textrm{PAF}_c(t)$ and $\textrm{PAF}_{LM}$. The estimates of $\textrm{PAF}_{crude}$ and $\textrm{PAF}_{LM}$ were obtained with generalized linear models using the R-function \textit{glm} \cite{stats2017}. Those for $\textrm{PAF}_o(t)$ and $\textrm{PAF}_c(t)$ were obtained with the R-function \textit{etm} \cite{etm2011}. To present the results we evaluated at each time point a summary (1st to 3rd quartile and the mean) of the 100 estimates.

\begin{table*}[htbp]
\begin{center}
\vspace{-3cm}
\begin{tabular}{cllllll}
\multicolumn{6}{c}{\large Data scenarios in the simulation study}\\
\hline
\hline
&&&&&\\
\multicolumn{6}{l}{ Scenarios with constant hazards} \\
\multicolumn{6}{l}{($\alpha_{ij}(t)=\alpha_{ij}$; $i=0,1$, $j=1,2,3,4,5$; cause specific hazard rates of the extended } \\
\multicolumn{6}{l}{illness-death model (Figure \ref{fig:extendedIllnessDeath})} \\
\hline
&&&&&\\
&\multicolumn{5}{c}{ Parameters of the cause-specific hazard rates}\\
&&&&&\\
 Scenario (fig. w/results)& \hspace{0.5cm} NI & Discharge  & Death & Discharge & Death\\
&&w/out NI&w/out NI&w/NI&w/NI\\
   \multicolumn{3}{l}{No effect on mortality:} &&&\\
 1 (fig. A1 (ESM)) &$\alpha_{01}=0.005$&$\alpha_{02}=0.02$&$\alpha_{03}=0.01$&$\alpha_{14}=0.02$&$\alpha_{15}=0.01$\\
  2 (fig. A2 (ESM))&$\alpha_{01}=0.05$&$\alpha_{02}=0.02$&$\alpha_{03}=0.01$&$\alpha_{14}=0.02$&$\alpha_{15}=0.01$\\
 &&&&&\\  
 \multicolumn{3}{l}{Direct effect on mortality:} &&&\\
  3 (fig. A3 (ESM))&$\alpha_{01}=0.005$&$\alpha_{02}=0.02$&$\alpha_{03}=0.01$&$\alpha_{14}=0.02$&$\alpha_{15}=0.02$\\
  4 (fig. \ref{fig:SimScene4})&$\alpha_{01}=0.05$&$\alpha_{02}=0.02$&$\alpha_{03}=0.01$&$\alpha_{14}=0.02$&$\alpha_{15}=0.02$\\
   &&&&&\\  
    \multicolumn{3}{l}{Indirect effect on mortality:} &&&\\
  5 (fig. A4 (ESM)) &$\alpha_{01}=0.005$&$\alpha_{02}=0.03$&$\alpha_{03}=0.01$&$\alpha_{14}=0.02$&$\alpha_{15}=0.01$\\
  6 (fig. A5 (ESM))&$\alpha_{01}=0.05$&$\alpha_{02}=0.03$&$\alpha_{03}=0.01$&$\alpha_{14}=0.02$&$\alpha_{15}=0.01$\\
\multicolumn{6}{c}{}\\
\hline
\multicolumn{6}{c}{}\\
\multicolumn{6}{l}{ Scenarios with time-varying Weibull hazards (fig. A6 and A8 (ESM))}\\
\multicolumn{6}{l}{($\alpha_{ij}(t)=k_{ij}\lambda_{ij}(\lambda_{ij}t)^{k_{ij}-1}$; $i=0,1$, $j=1,2,3,4,5$); cause specific hazard rates of the extended } \\
\multicolumn{6}{l}{illness-death model (Figure \ref{fig:extendedIllnessDeath})} \\
\hline
&&&&&\\
&\multicolumn{5}{c}{ Parameters of the cause-specific hazard rates}\\
&&&&&\\
 Scenario (fig. w/results)& \hspace{0.5cm} NI & Discharge  & Death & Discharge & Death\\
&&w/out NI&w/out NI&w/NI&w/NI\\
   \multicolumn{3}{l}{Indirect effect on mortality:} &&&\\
 7 (fig. \ref{fig:SimScene7}) &$k_{01}=1$&$k_{02}=1.4$&$k_{03}=0.9$&$k_{14}=1.4$&$k_{15}=0.9$\\
&$\lambda_{01}=0.06$ &$\lambda_{02}=0.08$&$\lambda_{03}=0.05$&$\lambda_{14}=0.05$&$\lambda_{15}=0.05$\\
 8 (fig. A7 (ESM)) &$k_{01}=1$&$k_{02}=0.9$&$k_{03}=1.4$&$k_{14}=0.9$&$k_{15}=1.4$\\
&$\lambda_{01}=0.06$ &$\lambda_{02}=0.08$&$\lambda_{03}=0.05$&$\lambda_{14}=0.05$&$\lambda_{15}=0.05$\\
 &&&&&\\  
   \multicolumn{3}{l}{Direct effect on mortality:} &&&\\
 9 (fig. A9 (ESM)) &$k_{01}=1$&$k_{02}=1.4$&$k_{03}=0.9$&$k_{14}=1.4$&$k_{15}=0.9$\\
&$\lambda_{01}=0.06$ &$\lambda_{02}=0.05$&$\lambda_{03}=0.05$&$\lambda_{14}=0.05$&$\lambda_{15}=0.08$\\
10 (fig. A10 (ESM)) &$k_{01}=1$&$k_{02}=0.9$&$k_{03}=1.4$&$k_{14}=0.9$&$k_{15}=1.4$\\
&$\lambda_{01}=0.06$ &$\lambda_{02}=0.05$&$\lambda_{03}=0.05$&$\lambda_{14}=0.05$&$\lambda_{15}=0.08$\\
\end{tabular}
\end{center}
\caption{\footnotesize Each scenarios was simulated 100 times with 10,000 and 2000 observations in each run. The R code was based on the simulation code provided by Heggland et al.\cite{heggland2015estimating}. In each run the four estimators of the PAF were calculated. Then, summary statistics over the 100 runs were obtained.}
\label{tab:scenarios}
\end{table*}

\subsection{Data settings with time-constant hazards}

\begin{figure*}[htb!]
\centering
\includegraphics[width=\textwidth]{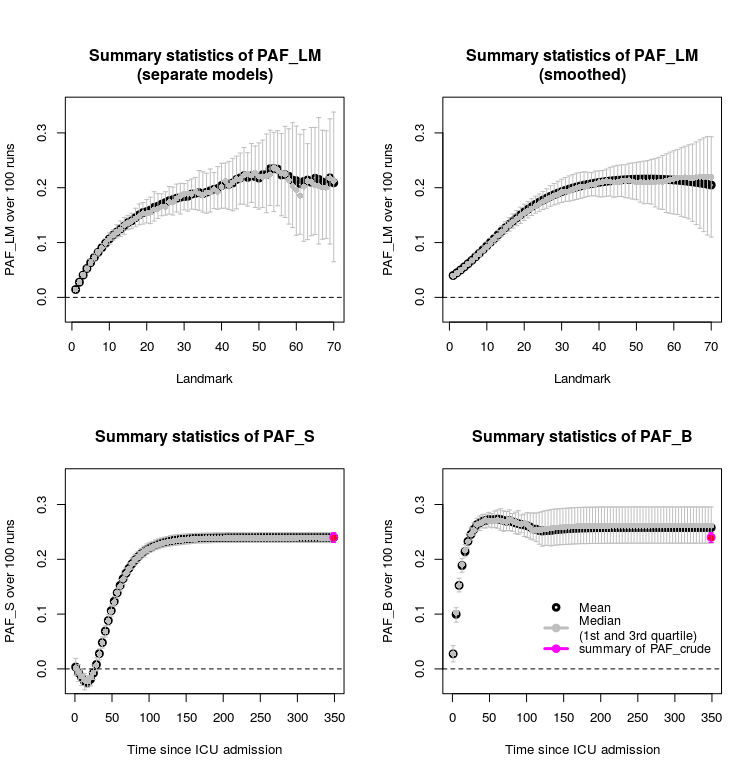}
\caption{Scenario 4: Simulation of a direct effect of the exposure on the death risk with constant cause-specific hazard rates ($\alpha_{01}(t)=0.05$, $\alpha_{02}(t)=0.02$, $\alpha_{03}(t)=0.01$, $\alpha_{14}(t)=0.02$ and $\alpha_{15}(t)=0.02$). Each sample consisted of 10,000 observations. 
The time window of $\widehat{PAF}_{LM}$ was the approximate mean length of stay (30 days).}
\label{fig:SimScene4}
\end{figure*}

We considered six different scenarios with constant hazards (see Table \ref{tab:scenarios}). In the following, we discuss Scenario 4 in more detail. In this scenario, we simulated a data setting with hazard rates $\alpha_{03}(t)=0.01$, $\alpha_{15}(t)=0.02$ and $\alpha_{02}(t)=\alpha_{14}(t)=0.02$. The infection hazard was quite high with $\alpha_{01}(t)=0.05$. Due to the inceased death hazard with infection, we expect an increased risk of death for exposed patients. The results of Scenario 4 (10,000 observations) are shown in Figure \ref{fig:SimScene4}. The results of Scenarios 1-3 and 5,6 lead to similar conclusions and are shown in the electronic supplementary material (ESM).

First, we consider $\widehat{\textrm{PAF}}_{LM}$ which is presented in the upper graphs of Figure \ref{fig:SimScene4} (separate models left, smoothed supermodel right). We find that due to an increasing prevalence of infection within the time window, $\widehat{\textrm{PAF}}_{LM}$ becomes higher at later landmarks. Thus, patients at risk at late time points would benefit most of a preventive intervention provided at the specific landmark. Regarding the estimators, the supermodel increases the efficiency of the separate models. However, variation of both estimators (separate models and supermodel) is quite large since the landmark datasets are much smaller in size than the initial population.

The $\textrm{PAF}_o(t)$ (lower graph on the left) is negative in the first part of the time frame. As explained previously \cite{schumacher2007attributable,von2017basic} this rather undesirable property results from death cases accumulating later among exposed patients than among unexposed patients. The apparent 'bump' resulting from this time delay is a characteristic of the \textit{estimand} and does not allow for a causal interpretation. The different simulation scenarios (see also the ESM) demonstrate that the infection hazard influences the depth of the bump and the speed of convergence of the estimand. 
Furthermore, the limit of $\textrm{PAF}_o(t)$ equals $\textrm{PAF}_{crude}$ which relates the observable proportion of death cases of eventually exposed and unexposed patients. This becomes not only apparent from the simulation study but can be also shown mathematically \cite{von2017basic}.
In the presence of censoring, $\widehat{\textrm{PAF}}_{crude}$ is unobservable and $\widehat{\textrm{PAF}}_o(\tau)$ is the best approximation of it. Finally, variation of $\widehat{\textrm{PAF}}_o(t)$ is minor and decreases with increasing infection hazard.

The $\widehat{\textrm{PAF}}_c(t)$ (lower graph, right) is positive at all time points and overcomes the interpretational limitations of $\textrm{PAF}_o(t)$ (see especially Scenarios 1 and 2 in the ESM).
In Scenario 4, $\widehat{\textrm{PAF}}_c(t)$ increases quickly and first exceeds $\widehat{\textrm{PAF}}_{crude}$ before decreasing again. Thus, after removal of the exposure risk deaths accumulate more slowly. This means that some death cases that could have been prevented around day 50 would still occur later in time. For constant hazards (however not generally \cite{vonCube2019beweis}) it can be shown mathematically that $\widehat{\textrm{PAF}}_c(t)$ should equal $\widehat{\textrm{PAF}}_{crude}$ at the end of follow-up. Nevertheless, despite complete follow-up, this is not the case in Scenario 4 due to the large infection hazard and the accordingly high systematic censoring of infected patients. Remark that the estimand $\textrm{PAF}_c(t)$ is independent of the infection hazard. However, as patients with an infection are treated as censored observations, the estimator $\widehat{\textrm{PAF}}_c(t)$ loses precision with an increasing infection hazard. Moreover, compared to $\widehat{\textrm{PAF}}_o(t)$ , $\widehat{\textrm{PAF}}_c(t)$ seems to be less precise. 

\subsection{Data settings with time-varying Weibull hazards}

\begin{figure*}[htb]
\centering
\includegraphics[width=\textwidth]{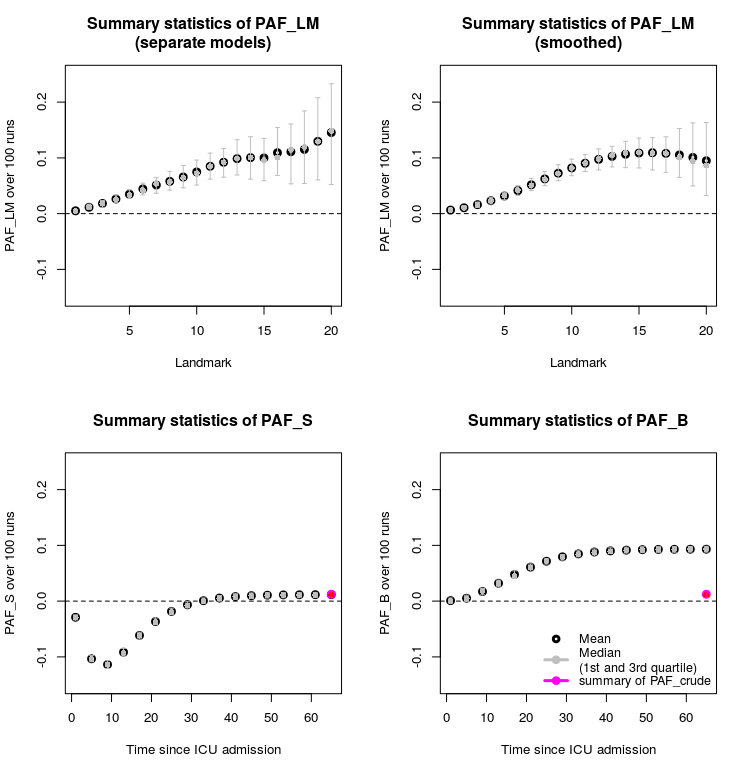}
\caption{Scenario 7: Simulation of an indirect effect of the exposure on the death risk with time-varying cause-specific Weibull hazard rates (the according hazards are shown in the upper graph of ESM Figure A6). Each sample consisted of 10,000 observations. 
The time window of $\widehat{PAF}_{LM}$ was the approximate mean length of stay (8 days).}
\label{fig:SimScene7}
\end{figure*}

To investigate the estimands and estimators of the PAF for a time-dependent exposure with time-varying hazards, we further simulated data based on Weibull hazards. We investigated four scenarios (see Table 1). In the following, we discuss Scenario 7 (Figure \ref{fig:SimScene7}, 10,000 observations) in detail. The argumentation and discussion are similar for the other scenarios (shown in the ESM; ESM Figures A7, A9 and A10). In Scenario 7, we assumed equal, decreasing death hazards and differential increasing discharge hazards. The discharge hazard of patients with infection was reduced compared to the one of patients without infection.

In Scenario 7, $\widehat{\textrm{PAF}}_{crude}$ is almost zero. This is rather surprising as we would expect an increased death risk due to a decreased discharge hazard with infection. The same applies for $\widehat{\textrm{PAF}}_o(t)$. It is negative in the first part of the time frame and then converges from below zero to $\widehat{\textrm{PAF}}_{crude}$. 
In contrast, both $\widehat{\textrm{PAF}}_c(t)$ and $\widehat{\textrm{PAF}}_{LM}$ imply an increased death risk for infected patients and a significant amount of preventable death cases at later time points. Due to the fact that there is no administrative censoring and no confounding, we are able to draw direct conclusions on the estimands.
This simulation setting shows that $PAF_{crude}$ has no causal interpretation \cite{vonCube2019beweis}.
The discrepancy between the results is mainly explained by the death hazards which are strongly decreasing within the first days. At this time, only a few patients already had an infection and most patients are at risk to die without an infection. Later, when the death hazards are already considerably decreased, more and more patients become at risk to die with infection. Therefore, the absolute number of patients that die without infection and with infection within the complete study period is almost the same despite an indirect effect of the infection via a decreased discharge hazard. However, interpretation must be done carefully as this equal number is rather due to the time delay of the occurrence of infections than the severity of infections. The time delay is not a causal consequence (though a natural aspect) of the exposure.

In the landmark approach patients that died or were discharged before the landmark no longer influence the estimand/estimate. The difference compared to $\textrm{PAF}_o(t)$ or $\textrm{PAF}_{crude}$ arises from considering different target populations and different unexposed patients. The $\widehat{\textrm{PAF}}_c(t)$ remains zero until the difference of the discharge hazards without and with NI becomes more pronounced. Then, $\widehat{\textrm{PAF}}_c(t)$ starts to increase and finally converges. The difference to $\textrm{PAF}_o(t)$ and $\textrm{PAF}_{crude}$ arises from the different definition of unexposed patients. 
The estimator $\widehat{\textrm{PAF}}_c(t)$ seems to give precise results as variation is small. However, due to the systematic censoring of infected patients, it can be unreliable. 
All estimators performed better when the sample size was 10,000. A sample size of 2,000 led to higher variations especially of the landmark approach. 
The main characteristics of the estimands and estimators of the PAF are summarized in Table 2.

\pagestyle{empty}
\begin{landscape}
\begin{table}[htbp]
\begin{center}
\begin{tabular}{l@{\hskip 0.5in}l@{\hskip 0.5in}l@{\hskip 0.35in}l}
\multicolumn{4}{c}{\large Estimands of the PAF for data with internal time-dependent exposure and competing risks}\\
\hline
\hline
\multicolumn{4}{c}{}\\
Definition & Time scale & Target population & Unexposed patients \\
\multicolumn{4}{c}{}\\
\hline
\multicolumn{4}{c}{}\\
$\textrm{PAF}_{crude}=\frac{P(D=1)-P(D=1|E=0)}{P(D=1)}$ &\begin{tikzpicture}
		\draw[|-|, thick] [draw=black](0,0)node[below]{0 } --(4,0); 
\draw [decorate,decoration={brace,amplitude=10pt},xshift=0pt,yshift=0pt, rotate=270]
(0,0) -- (0,4.0) node [black,midway,yshift=0.75cm] 
{\footnotesize complete study time};		
		\foreach \x  in {0}
  \draw[xshift=\x cm] (0pt,2pt) -- (0pt,-1pt) node[below,fill=white]{0};
  	\foreach \x  in {4}
  \draw[xshift=\x cm] (0pt,2pt) -- (0pt,-1pt) node[below,fill=white]{$\tau$};
	\end{tikzpicture} & All patients & Patients unexposed at
\\
\hspace{1.8cm}$=\textrm{PAF}_o(\tau)$& Summary measure && the end of their ICU-stay\\
&&&\\
&&&\\
$\textrm{PAF}_o(t)=\frac{P(D(t)=1)-P(D(t)=1|E(t)=0)}{P(D(t)=1)}$& 	\begin{tikzpicture}
		\draw[|->, thick] [draw=black](0,0)node[below]{0 } --node[below]{t (study time)}++(4,0); 
		\foreach \x  in {0}
  \draw[xshift=\x cm] (0pt,2pt) -- (0pt,-1pt) node[below,fill=white]{0};
	\end{tikzpicture}
& All patients & Patients unexposed at $t$\\
& Cumulative measure &&\\
&&&\\
&&&\\
$\textrm{PAF}_c(t)=\frac{P(D(t)=1)-P_0(D(t)=1)}{P(D(t)=1)}$&
	\begin{tikzpicture}
		\draw[|->, thick] [draw=black](0,0)node[below]{0 } --node[below]{t (study time)}++(4,0); 
		\foreach \x  in {0}
  \draw[xshift=\x cm] (0pt,2pt) -- (0pt,-1pt) node[below,fill=white]{0};
	\end{tikzpicture}
& All patients & All patients (hypo-
\\
& Cumulative measure && thetically unexposed)\\
&&&\\
&&&\\
$\textrm{PAF}_{LM}=\{\textrm{PAF}(l,h)|l\in \mathcal{LM}\}$& \begin{tikzpicture}
		
   \draw (0,0) -- (4,0);
    \foreach \x in {0,1,2.5,4}
      \draw (\x cm,3pt) -- (\x cm,-3pt);		
		\draw [decorate,decoration={brace,amplitude=10pt},xshift=0pt,yshift=0pt, rotate=270]
(0,1) -- (0,2) node [black,midway,yshift=0.5cm] 
{\footnotesize h};		
    \draw (0,0) node[below=3pt] {$ 0 $};
    \draw (1,0) node[below=3pt] {$ l_1 $};
    
    \draw (2.5,0) node[below=3pt] {$ l_2 $};
    \draw [decorate,decoration={brace,amplitude=10pt},xshift=0pt,yshift=0pt, rotate=270]
(0,2.5) -- (0,3.5) node [black,midway,yshift=0.5cm] 
{\footnotesize h};		
    \draw (4,0) node[below=3pt] {$\tau$};
  \end{tikzpicture}
  & All patients still & Patients unexposed \\
\hspace{2.2cm}$\textrm{PAF}(l,h)=P_{E_l}\times \frac{RR_{l,h}-1}{RR_{l,h}}$ & Summary measure &at risk at the LM&at the LM
 \\
\end{tabular}
\end{center}
\caption{\footnotesize D and D(t) are the variables of death at the end of the ICU stay and at time $t$ respectively; $P$ is the factual distribution of $D(t)$; $P_0$ the one after artificial removal of exposure risk; $P_{E_l}$ is the proportion of exposed at time $l$ among ICU death cases within $(l,l+h]$ of patients at risk at $l$; $RR_{l,h}$ is the relative risk of death in ICU within $(l, l+h]$ depending on exposure at landmark $l$ among patients at risk at $l$. $\mathcal{LM}$ is the set of landmarks.}
\end{table} 
\label{tab:resultsSim}

\begin{table}[htbp]
\vspace{-2.5cm}
\begin{center}
\begin{tabular}{l@{\hskip 0.5in}l@{\hskip 0.5in}l@{\hskip 0.35in}l}
\multicolumn{4}{c}{\large Estimands of the PAF for data with internal time-dependent exposure and competing risks}\\
\hline
\hline
\multicolumn{4}{c}{}\\
Estimand & Interpretation  & Advantages & Disadvantages  \\
&(see also Section 3)&&\\
\hline
\multicolumn{4}{c}{}\\
$PAF_{crude}$ & \% of observable attributable & Based on observable & Cannot capture time-varying\\
&  death cases at  & random variables & effects;\\
& the end of follow-up &&No causal interpretation\\
&&&\\
$PAF_o(t)$ & \% of observable attributable & Based on observable  & No causal interpretation\\
& death cases until $t$  & random variables & \\
&&&\\
$PAF_c(t)$ &\% of preventable death cases & Most consistent & \\
& until $t$ & with initial definition & \\
&&&\\
$PAF_{LM}$&\% of preventable death cases & Accounts for time dynamic& Depends on choice of LMs\\
& within time window $h$ & target population; & and time window;\\
& had the exposure been prevented  &Based on observable variables;& Does not result in a single number\\
&at the LM & Causal interpretation possible & \\
&&&\\
\hline
&&&\\
Estimator& R-function & Advantages & Disadvantages\\
&&&\\
\hline
&&&\\
$\widehat{\textrm{PAF}}_{crude}$ & \textit{glm} &  Easily adjusted for& No adjustment for\\
&&(time-independent) covariates;& time-dependent covariates\\
&& Efficient&\\
&&&\\
$\widehat{\textrm{PAF}}_o(t)$ &\textit{etm} &Efficient;& No adjustment for\\
&&& time-dependent covariates\\
&&&\\
$\widehat{\textrm{PAF}}_c(t)$ &\textit{etm} (Unadjusted; &Adjustable for& Adjustment for time-dependent cov. \\
&No R-functions for & time-dependent cov.&computer intensive and elaborate;\\
&adjusted analysis)&using inverse probability weights& Biased results if prevalence is high;\\
&&& Strong assumptions needed\\
&&&\\
$\widehat{\textrm{PAF}}_{LM}$ &\textit{glm}  &Easily adjusted for time-& Large sample size needed\\
&& dependent covariates&\\
\end{tabular}
\end{center}
\end{table} 
\end{landscape}
\pagestyle{plain}

\section{Data example: Preventable death cases among ventilated patients in intensive care}

In this section, we estimate the health impact of ventilator-associated pneumonia (VAP) for patients in intensive care. To do so, the four estimands of the PAF are derived and compared for a sample of the large French database Rea-Raisin (R\'{e}seau d'Alerte, d'Investigation et de Surveillance des Infections Nosocomiales). The data was collected from 2004 to 2015. The data sample includes information on 79,347 invasive-mechanically ventilated patients from 188 ICUs. All of these patients were at least two days ventilated with a mean length of stay (LOS) from first ventilation to discharge (dead or alive) of 17 days and a median of 12 days. After 21 days two third of the initial patient population has left the ICU. The shortest follow-up time was three days and the longest 403 days. A VAP was acquired by 8,320 patients of whom 2,746 died. Of the 71,027 patients that remained VAP-free until the end of their ICU stay 22,203 patients died. 

\begin{figure*}
\centering
\includegraphics[width=\textwidth]{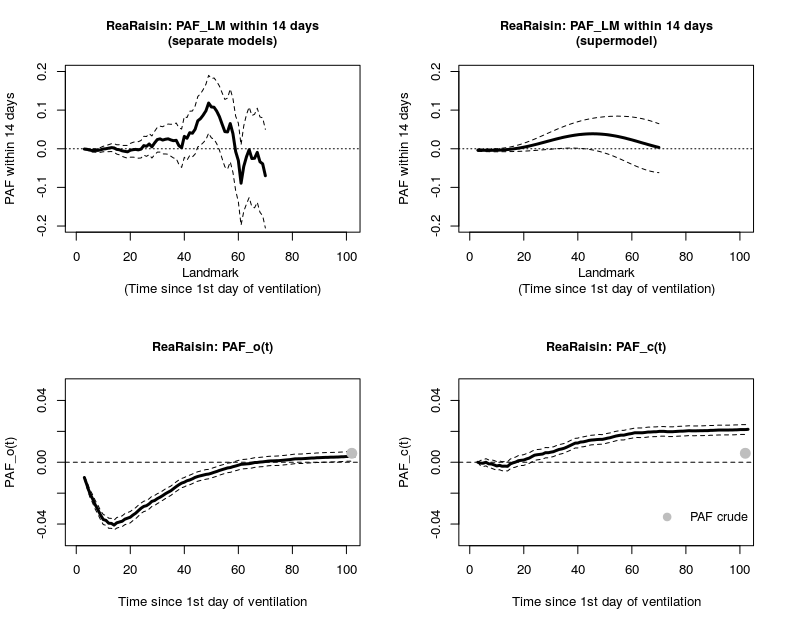}
\caption{Four estimates ($\widehat{PAF}_{LM}$ (separate models and supermodel), $\widehat{PAF}_o(t)$, $\widehat{PAF}_c(t)$ and $\widehat{PAF}_{crude}$) of the PAF of ICU mortality due to VAP for a sample of the Rea-Raisin database.}
\label{fig:ReaR_PAFs}
\end{figure*}

We estimate the PAF using the four discussed methods. For the landmark approach we use a time window of 14 days and landmarks daily from day 3 until day 70. Due to the definition of VAP,
day three is the first time point where a patient may die/be discharged with VAP. Moreover, only a small part of the ventilated patients stay in the ICU for more than 70 days.
The time window was chosen to be between the mean and the median LOS. The landmarks are such that enough events are observable in both patient groups (exposed and unexposed). The results are presented in Figure \ref{fig:ReaR_PAFs}.

The $\widehat{\textrm{PAF}}_{crude}$ was approximately 0.6\% which is the observable proportion of attributable cases due to VAP at the end of follow-up (after 403 days). The total number of attributable cases was 145. The $\widehat{\textrm{PAF}}_o(t)$ equals $\widehat{\textrm{PAF}}_{crude}$ at the end of follow-up. In contrast, $\widehat{PAF}_c(t)$ being interpretable as the proportion of preventable death cases was clearly larger with 2.22\% at the end of follow-up. This corresponds to a total number of 555 preventable ICU death cases.

Both, $\textrm{PAF}_o(t)$ and $\textrm{PAF}_c(t)$, describe how the proportion of attributable or respectively preventable cases accumulate over the course of time.
As already observed in the simulation study, $\widehat{PAF}_o(t)$ was negative in the first days. As commonly observed in data on patients in intensive care, the discharge hazard among uninfected patients is large in the beginning. Thus, the number of unexposed patients decreased quickly. The $\widehat{\textrm{PAF}}_c(t)$ was also negative in the first days but eventually converged to 0.022. As the infection hazard is low systematic censoring of infected patients does not affect the estimate $\widehat{\textrm{PAF}}_c(t)$ strongly.   

The $\widehat{\textrm{PAF}}_{LM}$ (separate models) was close to zero from the first landmarks until landmark 40. Then, it increased strongly reaching a peak of 11.8\% at day 50. Thus, if VAP could've been prevented at day 50 for patients still in the ICU at that day, then 11.8\% of the death cases occurring within the next 14 days could've been prevented. This corresponds to a total number of 50 preventable ICU death cases among these patients. The $\textrm{PAF}_{LM}$ demonstrates at which time points the health impact of VAP is the strongest. In this example, the smoothed $\widehat{\textrm{PAF}}_{LM}$ leads to more efficient but less precise results. The data example demonstrates that the supermodel should be complemented by the separate models. It is a weighted average of the separate models with more weight on early landmarks with many observations.
A more detailed discussion on the supermodel and the separate models as well as alternative smoothing methods is discussed in \cite{vonCube2019lm}.

To account for potential differences between exposed and unexposed patients, we adjusted for the available baseline covariates age, gender and severity of illness score at first day of ventilation. The estimator $\widehat{\textrm{PAF}}_{LM}$ was adjusted as described in Section 4 assuming a common RR among different patient groups. Both $\widehat{\textrm{PAF}}_o(t)$ and $\widehat{\textrm{PAF}}_c(t)$ were adjusted using the weighted average based on a Cox proportional hazards model as described in \cite{coeurjolly2012attributable}. The results differed only slightly from the unadjusted analysis. 

\section{Discussion}

In this article, we provide a comprehensive investigation of the PAF for time-dependent exposures and competing outcomes. By reviewing existing literature on the PAF and evaluating a novel estimand, we defined, identified and estimated the PAF in four different ways. Based on a theoretical exploration and with a simulation study, we discussed the differences in interpretation of the estimands, the advantages, and disadvantages of each approach as well as the performance of the corresponding estimators. The four estimands were used to quantify the burden of VAP for ventilated patients in intensive care on a population level.

The PAF as initially proposed by Levin \cite{Levin1953PAF} defines the attributable risk due to an exposure and is -- after sufficient control for confounding -- interpretable as proportion of preventable cases. In contrast, if the exposure depends on time, the effect measures $\textrm{PAF}_{crude}$ and $\textrm{PAF}_o(t)$, which are defined with conditional probabilities based on observable random variables, have no such causal interpretation. Nevertheless, they indicate how many of the observed cases are explained by the exposure.

The $\textrm{PAF}_c(t)$ seems to be the more natural extension of the PAF to time-dependent exposures. It is defined with hypothetical probabilities and is -- after control for confounding -- interpretable as proportion of preventable cases if exposure was eliminated. However, $\textrm{PAF}_c(t)$ is based on unobservable variables complicating estimation. Moreover, it fails to describe the \textit{observable} accumulation of death cases. In contrast, the $\textrm{PAF}_{LM}$ can be interpreted as both the 
population attributable risk and -- given sufficient adjustment for confounding and the Markov property -- the percentage of preventable cases.

In hospital-epidemiology, the commonly used estimand of the PAF due to an internal time-dependent exposure is $\textrm{PAF}_{crude}$. This estimand does not capture the temporal dynamics. The $\textrm{PAF}_o(t)$ is the time-dependent counterpart of $\textrm{PAF}_{crude}$. It is a more detailed description of how deaths accumulate over the course of time. At the end of follow-up in the absence of censoring $\textrm{PAF}_o(t)$ is equal to $\textrm{PAF}_{crude}$. 
In a simulation study, we demonstrated that the estimands $\textrm{PAF}_o(t)$ and $\textrm{PAF}_{crude}$ can differ substantially from the causal estimand $\textrm{PAF}_c(t)$. However, under constant hazards both $\textrm{PAF}_o(t)$ and $\textrm{PAF}_c(t)$ converge to $\textrm{PAF}_{crude}$. In this specific data situation, the $\textrm{PAF}_{crude}$ can be interpreted causally as the proportion of preventable cases. The simulation study showed that due to the systematic censoring of exposed patients in the estimation procedure, estimators of $\textrm{PAF}_c(t)$ can be inaccurate if the prevalence of the exposure is high. If the exposure prevalence is rather low, as in our data example, the bias is negligible. Moreover, the difference between $\textrm{PAF}_c(t)$ and $\textrm{PAF}_o(t)$ is small in such data settings.

Interpretation of $\textrm{PAF}_{LM}$ is substantially different from that of the other estimands. Being based on patients still at risk at specific time points, the percentage of preventable death cases is in reference to a time-dependent subpopulation of the initial population. The subpopulations allow to differentiate between short stayers and long stayers that are due to their extended length of stay at higher risk of NI acquisition \cite{wolkewitz2017landmark}. In contrast, $\textrm{PAF}_o(t)$, $\textrm{PAF}_c(t)$ and $\textrm{PAF}_{crude}$ refer to the initial population. 
While estimation of $\textrm{PAF}_{LM}$ and adjustment for (time-varying) confounding is straightforward, a clear drawback of this approach is the loss of precision due to the smaller sample size at the landmarks.

Finally, the choice of estimand depends on the intention of the study. To quantify the benefit of an intervention $\textrm{PAF}_c(t)$ and $\textrm{PAF}_{LM}$ are adequate measures. If the observable burden of an exposure on the population level is of main interest $\textrm{PAF}_o(t)$, $\textrm{PAF}_{crude}$ and $\textrm{PAF}_{LM}$ provide helpful insights. Decision on the estimand may be also data-driven. If the time of exposure is not available, only $\textrm{PAF}_{crude}$ can be estimated. If the no unmeasured confounding assumption is not reasonable, $\textrm{PAF}_o(t)$ and $\textrm{PAF}_{LM}$ are the preferred estimands.
We emphasize that care must be taken in the interpretation of the various estimands.

\subsection*{Acknowledgement}
The authors thank the 201 participating intensive care unit and infection control teams of the Rea-Raisin network. The Rea-Raisin network is supported by a grant from the French National Public Health Agency (Santé publique France).

\subsection*{Funding}
MvC and JFT were supported by the Innovative Medicines Initiative Joint Undertaking under grant agreement n [115737-2 – COMBACTE-MAGNET], resources of which are composed of financial contribution from the European Union’s Seventh Framework Programme (FP7/2007-2013) and EFPIA companies; MW has received funding from the German Research Foundation (Deutsche Forschungsgemeinschaft) under grant no. WO 1746/1-2.

\subsection*{Potential conflicts of interest}
Conflicts of interest for all authors: none.

\subsection*{Data availability}
The R code of the simulation study is provided as ESM.

\subsection*{Abbreviations}
PAF: Population-attributable fraction\\
ICU: Intensive-care unit\\
NI: Nosocomial infection

{\footnotesize
\bibliography{databaseCOMBACTEMagnet}

\begin{thebibliography}{10}

\bibitem{laxminarayan2013antibiotic}
R.~Laxminarayan, A.~Duse, C.~Wattal, A.~K. Zaidi, H.~F. Wertheim, N.~Sumpradit,
  E.~Vlieghe, G.~L. Hara, I.~M. Gould, H.~Goossens, {\em et~al.}, ``Antibiotic
  resistance—the need for global solutions,'' {\em The Lancet {I}nfectious
  {D}iseases}, vol.~13, no.~12, pp.~1057--1098, 2013.

\bibitem{Levin1953PAF}
M.~L. Levin, ``The occurrence of lung cancer in man,'' {\em Acta Unio
  Internationalis contra Cancrum}, vol.~9, p.~531–541, 1953.

\bibitem{chen2006attributable}
Y.~Q. Chen, C.~Hu, and Y.~Wang, ``Attributable risk function in the
  proportional hazards model for censored time-to-event,'' {\em Biostatistics},
  vol.~7, no.~4, pp.~515--529, 2006.

\bibitem{chen2010attributable}
L.~Chen, D.~Lin, and D.~Zeng, ``Attributable fraction functions for censored
  event times,'' {\em Biometrika}, vol.~97, no.~3, pp.~713--726, 2010.

\bibitem{samuelsen2008attributable}
S.~O. Samuelsen and G.~E. Eide, ``Attributable fractions with survival data,''
  {\em Statistics in {M}edicine}, vol.~27, no.~9, pp.~1447--1467, 2008.

\bibitem{laaksonen2010piecewise}
M.~A. Laaksonen, P.~Knekt, T.~H{\"a}rk{\"a}nen, E.~Virtala, and H.~Oja,
  ``Estimation of the population attributable fraction for mortality in a
  cohort study using a piecewise constant hazards model,'' {\em American
  {J}ournal of {E}pidemiology}, vol.~171, no.~7, pp.~837--847, 2010.

\bibitem{laaksonen2010estimation}
M.~Laaksonen, T.~H{\"a}rk{\"a}nen, P.~Knekt, E.~Virtala, and H.~Oja,
  ``Estimation of population attributable fraction (paf) for disease occurrence
  in a cohort study design,'' {\em Statistics in {M}edicine}, vol.~29, no.~7-8,
  pp.~860--874, 2010.

\bibitem{sjolander2016cautionary}
A.~Sj{\"o}lander, ``A cautionary note on the use of attributable fractions in
  cohort studies,'' {\em Statistical {M}ethods in {M}edical {R}esearch},
  vol.~25, no.~6, pp.~2434--2443, 2016.

\bibitem{zhao2017onestimation}
W.~Zhao, Y.~Q. Chen, and L.~Hsucorresponding, ``On estimation of time-dependent
  attributable fraction from population-based case-control studies,'' {\em
  Biometrics}, vol.~73, no.~3, pp.~866--875, 2017.

\bibitem{sjolander2017doubly}
A.~Sj{\"o}lander and S.~Vansteelandt, ``Doubly robust estimation of
  attributable fractions in survival analysis,'' {\em Statistical {M}ethods in
  {M}edical {R}esearch}, vol.~26, no.~2, pp.~948--969, 2017.

\bibitem{wolkewitz2014interpreting}
M.~Wolkewitz, B.~S. Cooper, M.~J. Bonten, A.~G. Barnett, and M.~Schumacher,
  ``Interpreting and comparing risks in the presence of competing events,''
  {\em BMJ}, vol.~349, p.~g5060, 2014.

\bibitem{schumacher2013hospital}
M.~Schumacher, A.~Allignol, J.~Beyersmann, N.~Binder, and M.~Wolkewitz,
  ``Hospital-acquired infections -- appropriate statistical treatment is
  urgently needed!,'' {\em International {J}ournal of {E}pidemiology}, vol.~42,
  no.~5, pp.~1502--1508, 2013.

\bibitem{schumacher2007attributable}
M.~Schumacher, M.~Wangler, M.~Wolkewitz, and J.~Beyersmann, ``Attributable
  mortality due to nosocomial infections. a simple and useful application of
  multistate models.,'' {\em Methods of {I}nformation in {M}edicine}, vol.~46,
  no.~5, p.~595, 2007.

\bibitem{bekaert2010adjusting}
M.~Bekaert, S.~Vansteelandt, and K.~Mertens, ``Adjusting for time-varying
  confounding in the subdistribution analysis of a competing risk,'' {\em
  Lifetime {D}ata {A}nalysis}, vol.~16, no.~1, pp.~45--70, 2010.

\bibitem{vonCube2019beweis}
M.~von Cube, M.~Schumacher, and M.~Wolkewitz, ``Causal inference with
  multi-state models - estimands and estimators of the population-attributable
  fraction,'' Tech. Rep. arXiv:1903.10315, arXiv, March 2019.

\bibitem{vonCube2019lm}
M.~von Cube, M.~Schumacher, H.~Putter, J.-F. Timsit, C.~van~de Velde, and
  M.~Wolkewitz, ``The population-attributable fraction for time-dependent
  exposures using dynamic prediction and landmarking,'' Tech. Rep.
  arXiv:1904.07295, arXiv, April 2019.

\bibitem{van2011dynamic}
H.~van Houwelingen and H.~Putter, {\em Dynamic {P}rediction in {C}linical
  {S}urvival {A}nalysis}.
\newblock CRC Press, 2011.

\bibitem{benichou2001review}
J.~Benichou, ``A review of adjusted estimators of attributable risk,'' {\em
  Statistical {M}ethods in {M}edical {R}esearch}, vol.~10, no.~3, pp.~195--216,
  2001.

\bibitem{miettinen1974proportion}
O.~S. Miettinen, ``Proportion of disease caused or prevented by a given
  exposure, trait or intervention,'' {\em American {J}ournal of
  {E}pidemiology}, vol.~99, no.~5, pp.~325--332, 1974.

\bibitem{benichou2005attributable}
J.~Benichou, ``Attributable risk,'' {\em Encyclopedia of {B}iostatistics},
  2005.

\bibitem{dahlqwist2016model}
E.~Dahlqwist, J.~Zetterqvist, Y.~Pawitan, and A.~Sj{\"o}lander, ``Model-based
  estimation of the attributable fraction for cross-sectional, case--control
  and cohort studies using the r package af,'' {\em European {J}ournal of
  {E}pidemiology}, vol.~31, no.~6, pp.~575--582, 2016.

\bibitem{greenland1993maximum}
S.~Greenland and K.~Drescher, ``Maximum likelihood estimation of the
  attributable fraction from logistic models,'' {\em Biometrics}, pp.~865--872,
  1993.

\bibitem{beyersmann2011competing}
J.~Beyersmann, A.~Allignol, and M.~Schumacher, {\em Competing {R}isks and
  {M}ultistate {M}odels with R}.
\newblock Springer, 2011.

\bibitem{coeurjolly2012attributable}
J.-F. Coeurjolly, M.~Nguile-Makao, J.-F. Timsit, and B.~Liquet, ``Attributable
  risk estimation for adjusted disability multistate models: application to
  nosocomial infections,'' {\em Biometrical Journal}, vol.~54, no.~5,
  pp.~600--616, 2012.

\bibitem{eide2001attributable}
G.~E. Eide and I.~Heuch, ``Attributable fractions: fundamental concepts and
  their visualization,'' {\em Statistical Methods in Medical Research},
  vol.~10, no.~3, pp.~159--193, 2001.

\bibitem{robins2000marginal}
J.~M. Robins, M.~A. Hernan, and B.~Brumback, ``Marginal structural models and
  causal inference in epidemiology,'' 2000.

\bibitem{nicolaie2013dynamic}
M.~Nicolaie, J.~van Houwelingen, T.~de~Witte, and H.~Putter, ``Dynamic
  pseudo-observations: A robust approach to dynamic prediction in competing
  risks,'' {\em Biometrics}, vol.~69, no.~4, pp.~1043--1052, 2013.

\bibitem{greenland1987variance}
S.~Greenland, ``Variance estimators for attributable fraction estimates
  consistent in both large strata and sparse data,'' {\em Statistics in
  {M}edicine}, vol.~6, no.~6, pp.~701--708, 1987.

\bibitem{sjolander2011estimation}
A.~Sj{\"o}lander, ``Estimation of attributable fractions using inverse
  probability weighting,'' {\em Statistical {M}ethods in {M}edical {R}esearch},
  vol.~20, no.~4, pp.~415--428, 2011.

\bibitem{heggland2015estimating}
T.~Heggland, ``Estimating transition probabilities for the illness-death
  model,'' Master's thesis, UiO University of Oslo -- Faculty of Mathematics
  and Natural Sciences, 2015.

\bibitem{stats2017}
{R Core Team}, {\em R: A Language and Environment for Statistical Computing}.
\newblock R Foundation for Statistical Computing, Vienna, Austria, 2017.

\bibitem{etm2011}
A.~Allignol, M.~Schumacher, and J.~Beyersmann, ``Empirical transition matrix of
  multi-state models: The {etm} package,'' {\em Journal of Statistical
  Software}, vol.~38, no.~4, pp.~1--15, 2011.

\bibitem{von2017basic}
M.~von Cube, M.~Schumacher, and M.~Wolkewitz, ``Basic parametric analysis for a
  multi-state model in hospital epidemiology,'' {\em BMC Medical Research
  Methodology}, vol.~17, no.~1, p.~111, 2017.

\bibitem{wolkewitz2017landmark}
M.~Wolkewitz, M.~Zortel, M.~Palomar-Martinez, F.~Alvarez-Lerma,
  P.~Olaechea-Astigarraga, and M.~Schumacher, ``Landmark prediction of
  nosocomial infection risk to disentangle short-and long-stay patients,'' {\em
  Journal of Hospital Infection}, vol.~96, no.~1, pp.~81--84, 2017.

\bibitem{andersen2008inference}
P.~K. Andersen and M.~P. Perme, ``Inference for outcome probabilities in
  multi-state models,'' {\em Lifetime {D}ata {A}nalysis}, vol.~14, no.~4,
  pp.~405--431, 2008.

\bibitem{allignol2011estimating}
A.~Allignol, M.~Schumacher, and J.~Beyersmann, ``Estimating summary functionals
  in multistate models with an application to hospital infection data,'' {\em
  Computational {S}tatistics}, vol.~26, no.~2, pp.~181--197, 2011.

\end{thebibliography}
\bibliographystyle{ieeetr}

}

\section*{Appendix A: Mathematical identification of $\textrm{PAF}_c(t)$}

In this section we identify the estimand
\begin{equation*}
\textrm{PAF}_c(t)=\frac{P(D(t)=1)-P_0(D(t)=1)}{P(D(t)=1)}
\end{equation*}
mathematically. To do so, we consider two distributions, $P$ and $P_0$, of the indicator $D(t)$. We defined $D(t)$ in Section 2 by $D(t)=1$ if the patient died in the ICU by time $t$, otherwise $D(t)=0$.
$P$ corresponds to the distribution of $D(t)$ in the real world and $P_0$ to that in a hypothetical world where infections could be prevented. An identification of the PAF (for cohort studies of fixed length) with this approach was introduced by Eide et al.\cite{eide2001attributable}.

\begin{figure*}[htb!]
\centering
\includegraphics[scale=1]{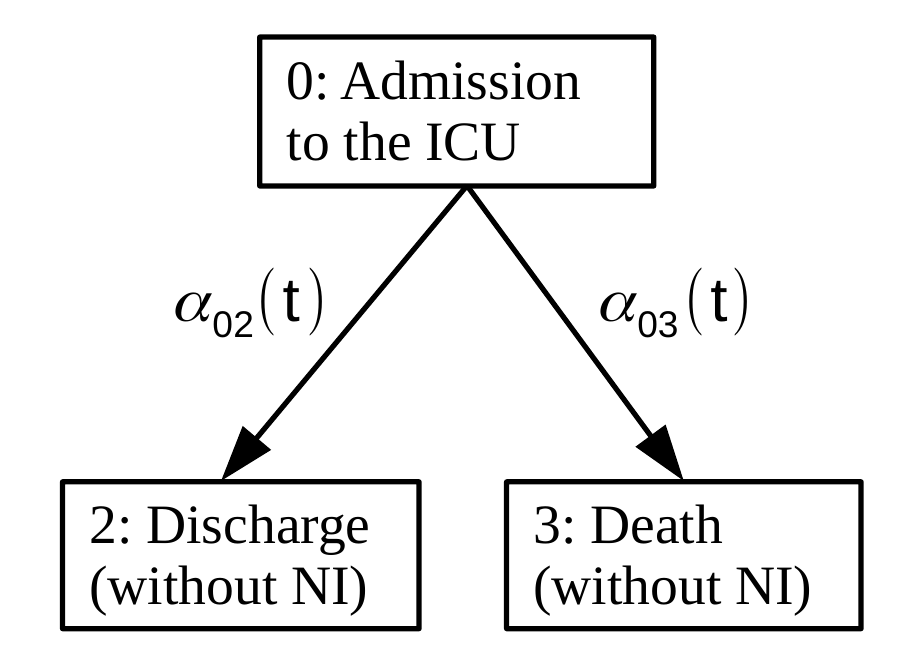}
 \caption{Competing risks model with hazard rates $\alpha_{02_{0}}(t)=\alpha_{02}(t)$ and $\alpha_{03_{0}}(t)=\alpha_{03}(t)$.}
  \label{fig:CR_model}
\end{figure*}

To identify the two distributions, we consider two counting processes. These are based on the extended illness-death model shown in Figure \ref{fig:extendedIllnessDeath} and the competing risks model shown in Figure \ref{fig:CR_model}. The counting process based on the extended illness-death model is defined by $(X(t), 0\leq{t}\leq{\tau})$ and has the finite state space $S=\bigl\{0,1,2, 3,4,5 \bigr\}$.
The transition probabilities are defined by
\begin{equation*}
P_{ij}(s,t)=P(X(t)=j|X(s)=i, \mathscr{X}_{s^-}), 
\end{equation*}
where $0\leq{s}\leq{t}\leq{\tau}$, and $i,j\in{S}$.
The stochastic process is fully characterized by the cause-specific transition hazards -- the instantaneous risk of moving from one state to another.
These quantities are defined as
\begin{equation*}
\alpha_{ij}(t)=\lim\limits_{dt\to 0}\frac{P_{ij}(t,t+dt)}{dt},
\end{equation*}
where $0\leq{t}\leq{\tau},\quad i,j\in{S}$. For more details we refer to Andersen et al.\cite{andersen2008inference}.
The counting process $\mathscr{X}$ describes the distribution $P$ of $D(t)$ in the real world where patients may acquire an infection. The possibility to acquire an infection is modelled by State 1 and the infection hazard $\alpha_{01}(t)$. The observable death risk in the real world was identified in Section 2 in \eqref{P_Dt}.

The counting process based on the competing risks model is defined by $(Y(t), 0\leq{t}\leq{\tau})$ with finite state space $R=\bigl\{0,2, 3 \bigr\}$. The end of study time is denoted by $\tau$ and is in our situation the same as (or smaller than) that of $\mathscr{X}$.
The transition probabilities of counting process $\mathscr{Y}$ with history $\mathscr{Y}_s$ are given by
\begin{equation*}
P_{ij_{0}}(s,t)=P_0(Y(t)=j|Y(s)=i, \mathscr{Y}_{s^-}), 
\end{equation*}
where $0\leq{s}\leq{t}\leq{\tau}$, and $i,j\in{R}$. Moreover, the hazard rates are defined by
\begin{equation*}
\alpha_{ij_{0}}(t)=\lim\limits_{dt\to 0}\frac{P_{ij_{0}}(t,t+dt)}{dt},
\end{equation*}
where $0\leq{t}\leq{\tau},\quad i,j\in{R}$.

The counting process $\mathscr{Y}$ describes the hypothetical world, where everything remains the same as in the real world except for the fact that infections can be prevented. This implies that the rates of moving to State 2 and 3 in the competing risks model are the same as in the extended illness-death model i.e. $\alpha_{02_{0}}(t)=\alpha_{02}(t)$ and $\alpha_{03_{0}}(t)=\alpha_{03}(t)$. However, the transition probabilities of the two counting processes differ. This is due to the fact that the infection hazard $\alpha_{01}(t)$ of the extended illness-death model is generally greater than zero. Only, in the special case where $\alpha_{01}(t)=0$ do we have $P_{ij}(s,t)=P_{ij_{0}}(s,t)$ for those $i$ and $j$ where both quantities are defined (i.e. $i=0$ and $j=2,3$).

Similarly to Allignol et al.\cite{allignol2011estimating}, we define $T:=\inf\{t>0| Y(t)\in \{2,3\}\}$ as the final event time and $Y_{T}=Y(T)$ as the final absorbing state for the competing risks model.
Since under counting process $\mathscr{Y}$, $D(t)=\mathbbm{1}_{T\le t, X_T=3}$ we identify the probability of $D(t)$ in the hypothetical world by
\begin{equation}
P_0(D(t)=1)=P_0(T\le t, X_T=3)=P_{03_0}(0,t).
\label{P0_Dtestimand}
\end{equation}

\end{document}